\documentclass[runningheads]{llncs}
\usepackage{graphicx}
\begin{document}
\title{\textit{MAM-E}: Mammographic synthetic image generation with diffusion models}
%
%

\author{Ricardo Montoya-del-Angel\inst{1}, Karla Sam-Millan\inst{1}, Joan C Vilanova\inst{2}
\and Robert Martí\inst{1}  
}

%
\authorrunning{Montoya-del-Angel et al.}

\institute{Computer Vision and Robotics Institute (ViCOROB), University of Girona, Spain \and Department of Radiology, Clínica Girona, Institute of Diagnostic Imaging (IDI) Girona,
University of Girona, Spain\\
\email{ricardo.montoya@udg.edu}
}

%
\maketitle              
\begin{abstract}
Generative models are used as an alternative data augmentation technique to alleviate the data scarcity problem faced in the medical imaging field. Diffusion models have gathered special attention due to their innovative generation approach, the high quality of the generated images and their relatively less complex training process compared with Generative Adversarial Networks. Still, the implementation of such models in the medical domain remains at early stages. In this work, we propose exploring the use of diffusion models for the generation of high quality full-field digital mammograms using state-of-the-art conditional diffusion pipelines. Additionally, we propose using stable diffusion models for the inpainting of synthetic lesions on healthy mammograms. We introduce \textit{MAM-E}, a pipeline of generative models for high quality mammography synthesis controlled by a text prompt and capable of generating synthetic lesions on specific regions of the breast. Finally, we provide quantitative and qualitative assessment of the generated images and easy-to-use graphical user interfaces for mammography synthesis.

\end{abstract}
\section{Introduction}

Data scarcity is an important problem faced in the medical imaging domain, caused by several factors such as expensive image acquisition, processing and labeling procedure, data privacy concerns, and the rare incidence of some pathologies \cite{kazerouni_diffusion_2023}. This leads a reduction of the volume of medical data available for the training of deep learning models, which limits the models performance and holds back the development of computer-aided systems, compared with non-medical imaging applications.\par

Generative models have been used to complement traditional data augmentation techniques and expand medical datasets, with Generative Adversarial Networks (GANs) being the state-of-the-art (SOTA) due to their high image quality and photorealism. Nevertheless, unstable training, low diversity generation and low sample quality make the use of GAN-like architectures challenging for medical data \cite{kazerouni_diffusion_2023}, as medical diagnosis can depend on subtle changes in the organs appearance reflected in the images, affecting the performance of a computer-assisted diagnosis and intervention systems \cite{muller-franzes_diffusion_2022}.

Diffusion models (DM) captured special attention from the generative models community when they were proposed for image generation and seemingly outperformed GANs in 2021 \cite{dhariwal_diffusion_2021}. Since then applications and research papers for medical images have been published to explore this new image generation principle. For instance, \cite{dorjsembe_three-dimensional_2022} proposed using the original pipeline of diffusion models on computer vision called denoising diffusion probabilistic models (DDPM) \cite{ho_denoising_2020} for the generation of high-quality MRI of brain tumors. This first implementation of diffusion models for 3D medical images reached SOTA results and outperformed the baseline models based on 3D GANs. Latent diffusion was used by \cite{pinaya_brain_2022} to generate high-resolution 3D brain images, increasing the image resolution from 64x64x64 to 160x224x160 without requiring more GPU memory usage or overall training time. The Fréchet Inception Distance (FID) for image fidelity, and the multi-scale structural similarity index measure (MS-SSIM) for generation diversity were computed and in both cases DM surpassed the GANs baseline metrics. \par

A Stable Diffusion (SD) implementation for medical images was introduced by \cite{chambon_roentgen_2022} who proposed a model for chest X-ray generation. Their model, named \textit{RoentGen}, was able to create visually convincing, diverse chest X-rays, controlling the generation output using text prompts with radiology-specific language. A key characteristic of this work is the use of SD weights pretrained with natural images as baseline. Instead of training from scratch specific parts of the network were fine-tuned to adapt the weights from its original to a new medical domain. This DM fine-tuning approach is called \textit{Dreambooth} and was first introduced by \cite{ruiz_dreambooth_2023} for natural images. \par

Besides full-field image generation, DM can be used for other tasks such as image inpainting. Some works have explored lesion inpainting using DM for brain MRI such as \cite{rouzrokh_multitask_2022} from the Mayo Clinic. They developed a DDPM to execute several inpainting tasks, like generating synthetic lesions or healthy tissue, on slices of the 3D volumes in various sequences. Their model was capable of generate realistic tumoral lesions and tumor-free brain tissue, although the performance of the model was only assessed visually.\par

\begin{figure}
\includegraphics[width=\textwidth]{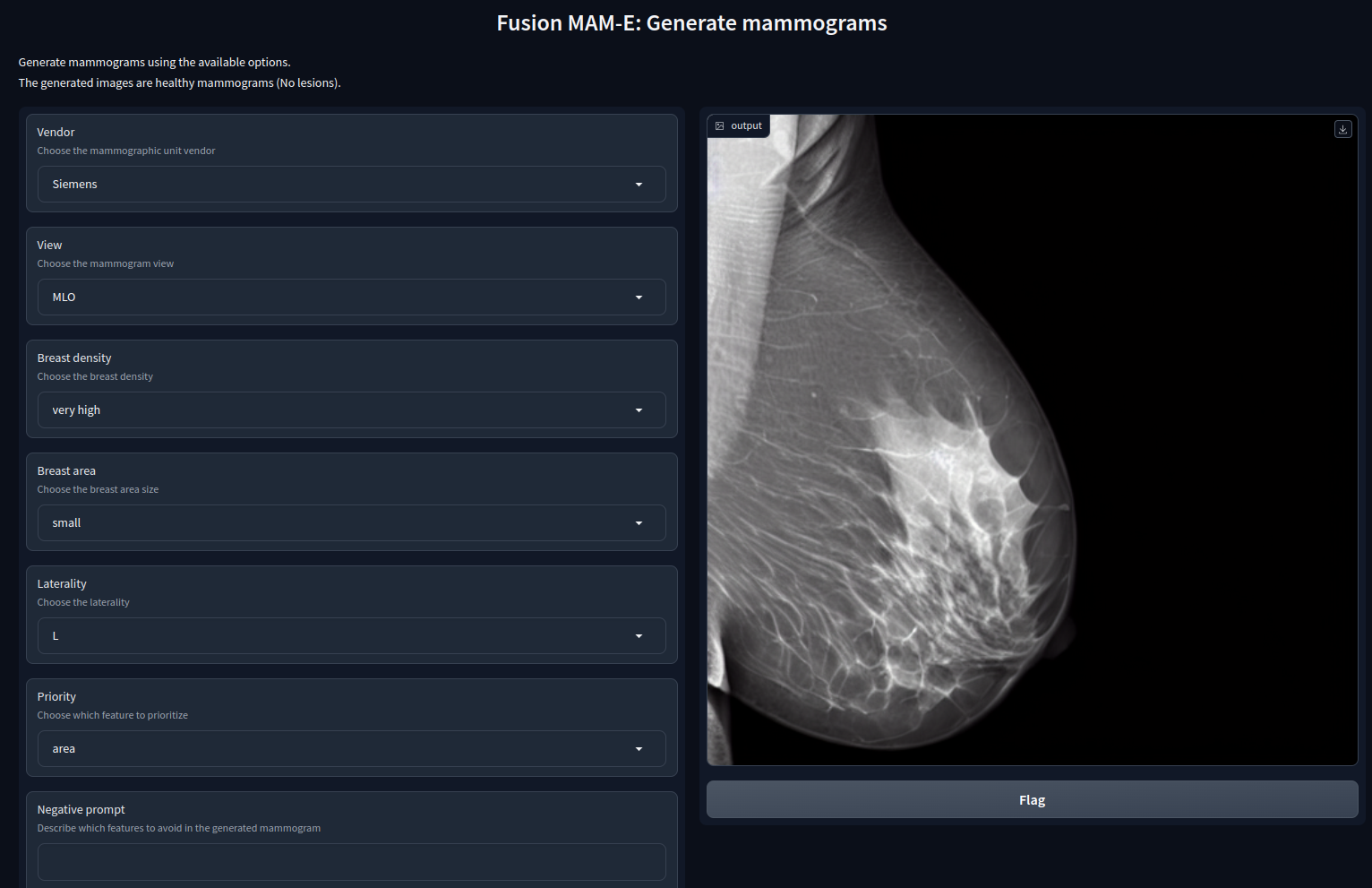}
\caption{Graphical user interface of \textit{MAM-E} for generation of synthetic healthy mammograms} \label{fig:mame_example}
\end{figure}

Despite all this, the use of diffusion models in the medical imaging field continues at early stages, specially for mammography. Prior to this publication we have released the source code, weights and user interface for the first implementation of SD for mammographic image synthesis in \cite{montoya-del-angel_mam-e_nodate}. Following works have explored the generation of synthetic mammograms using DM, such as the release of one synthetic mammography dataset from \cite{pinaya_generative_2023}, composed of 100k 512x512 synthetic images with masking level labeling, and the proposal of \cite{kidder_advanced_2023} to explore the use of SD for brain imaging and contrast-enhanced spectral mammography.\par

We introduce \textit{MAM-E}, a pipeline of generative models for high quality mammographic image synthesis, capable of generating images based on a text prompt description, and also capable of generating lesions on a specific section of the breast using a mask. Our pipeline was developed using stable diffusion, a SOTA diffusion model technique that uses both conditioning, to control the image generation, and a latent space to allow high-resolution without requiring large computational resources. The generated images are \textit{for presentation}, meaning that their appearance and pixel intensities are meant for radiologist inspection, with the limitations on resolution and pixel depth inherent to the current state of diffusion pipelines. To the knowledge of the authors, this is the first work to use stable diffusion fine-tuning for lesion inpainting for mammography. Moreover, this work source code publication represented the first implementation of SD for mammographic image generation. \par

Our main pipeline can be separated into two tasks: healthy mammogram generation and lesion inpainting. For the first task, the generation process is controlled by a text conditioning with the description of the image including view, breast density, breast area and vendor. For the second task we use an stable diffusion inpainting model designed to generate synthetic lesions in desired regions of the a mammogram. The name of our model was inspired by OpenAI's DALL-E \cite{ramesh_zero-shot_2021}. The source code (https://github.com/Likalto4/diffusion-models\_master) and the pretrained weights (https://huggingface.co/Likalto4) are publicly available. Additionally, graphical user interfaces for both synthesis tasks were designed for easy-to-use image generation and their source code can be found in the same repository with the characteristics shown in figure \ref{fig:mame_example}.\par

\section{Methodology}

\subsection{Datasets}

We decided to use two datasets for the training of the diffusion models so that different patient populations and mammography unit vendors were considered. \par

\subsubsection{OMI-H} We used a subset of the OPTIMAM Mammography Image Database, consisting of around 40k Hologic vendor full-field digital mammograms (FFDM) from several UK breast screening centers and with different image views \cite{halling-brown_optimam_2021}. The dataset was composed of images with and without lesions (benign, malignant and interval-cancers), and expert annotations are included in the respective cases, including the coordinates of a bounding box surrounding the lesion.

\subsubsection{VinDr-Mammo} A second dataset composed of around 20k FFDM with breast-level assessment and extensive lesion annotation was also used. It consists of 5,000 mammography exams, each with 4 standard views (CC and MLO for both lateralities), coming from two primary hospitals from Vietnam, giving a total of 20,000 images in DICOM files \cite{nguyen_vindr-mammo_2023}. Metadata of each image consisting of both technical and clinical information waas also available in a CSV file. We filter the images so that only mammograms coming from a Siemens vendor unit were used.

Table \ref{table:dataset} shows the distribution of the cases among both datasets and their combination.

\begin{table}
\caption{Distribution of cases for both datasets.}
\centering
\begin{tabular}{lrrrl}
\hline
                                   & OMI-H &  VinDr & Combined &  \\ \hline
Healthy     & 33,643                    & 13,942                                           & 47,585                       &  \\
With lesion & 6,908                                            & 1,533                                            & 8,441                        &  \\
 Total       & 40,551                                           & 15,475                                           & 56,026                       &  \\ \hline
\end{tabular}
\label{table:dataset}
\end{table}

\subsection{Data preprocessing and preparation}

Both datasets were subject to the same preprocessing and preparation steps. First, mammograms were saved as PNG files to ensure faster access and less disk memory space. Secondly, to be able to use pretrained weights, the images were saved in RGB format, repeating the original gray-channel into each RGB channel. The original image intensities with \textit{uint16} data type were scaled to a [0, 255] range with a reduced \textit{uint8} data type.\par

\subsubsection{Healthy image generation} For each healthy mammogram a text prompt description was created and saved along with the image ID in a JSON file. In the case of the OMI-H dataset we created a prompt with the image view and breast area size information. We defined a criterion to categorize the breast area sizes in three main groups: small, medium and large. For the VinDr dataset the breast density information was included instead of the breast area for the prompt description. Breast density was available in BI-RADS scale so we needed to transform this information into a semantically meaningful text. The criteria used for both cases is defined in table \ref{fig:criteria}.\par

\begin{table}[!t]
\caption{Criteria for breast area size and breast density.}
\begin{minipage}{0.5\textwidth}
\begin{center}

\begin{tabular}{ll}
        \hline
        \multicolumn{2}{c}{{Breast area size}}                              \\ \hline
        { Small}  & { ratio \textless 0.4}               \\
        { Medium} & { 0.4 \textless ratio \textless 0.6} \\
        { Big}    & { ratio \textgreater 0.6}            \\ \hline
\end{tabular}

\end{center}
\end{minipage}
\begin{minipage}{0.5\textwidth}
\begin{center}

\begin{tabular}{ll}
        \hline
        \multicolumn{2}{c}{{ Breast density}}           \\ \hline
        { Very low}  & { Density A} \\
        { Low}       & { Density B} \\
        { High} & { Density C} \\
        { Very high}      & { Density D} \\ \hline
\end{tabular}

\end{center}
\end{minipage}
\label{fig:criteria}
\end{table}

\subsubsection{Lesion inpainting} The inpainting task requires mammograms with confirmed lesions only. Using the bounding boxes coordinates available in the metadata, binary masks were generated. Naturally, due to the resizing and cropping preprocessing performed previously, the original coordinates required a proper redefinition using simple geometrical properties. The mask has pixel values of $255$ inside the bounding box and zero elsewhere. Because the SD architecture used for the inpainting task requires an input text prompt for the generation, a toy prompt with "a mammogram with a lesion" text was used for all training images.\par

\subsection{Diffusion models}

The original diffusion model idea was presented by \cite{sohl-dickstein_deep_2015} and consisted on using a Markov chain, a sequence of stochastic events whose time steps depend on the previous one, to gradually convert one known distribution (e.g. Gaussian distribution) into another (target distribution). Inspired by non-equilibrium statistical physics, the main idea is to systematically and iteratively destroy structure in a data distribution through a process called \textbf{forward diffusion}. Then, the \textbf{reverse diffusion process} is learned and used to restore structure in data. The first practical implementation of the DM premise on images was developed by \cite{ho_denoising_2020} introducing \textit{Denoising diffusion probabilistic models} (DDPM). In this framework, the data is destroyed by adding Gaussian noise to the image in an iterative fashion described by the Markov chain. The total number of diffusion timesteps $T$ is defined by the user but an usual number is around $T=1000$. To learn the reverse process a UNet is used to carry on the denoising process.\par

To solve the image size limitation, latent diffusion was introduced, which uses encoders to \textit{compress} images from their original sizes in the image space into a smaller representation in the latent space. The motivation behind this is that images usually contain redundant information and an encoder can produce a smaller representation that can later be reconstructed back using a decoder. Therefore, in latent diffusion the diffusion processed is performed on the latent representations rather than the original images \cite{rombach_high-resolution_2022}.\par

Stable diffusion is an improvement to \cite{rombach_high-resolution_2022} latent diffusion work, in which text conditioning is added to the model for additional control on the generation process. The text conditioning is a prompt with the description of the image. To create a numeric representation of the prompt a pretrained transformer called CLIP is used \cite{radford_learning_2021}. CLIP, which stands for Contrastive Language-Image Pre-training, maps both text and images into the same representational space, allowing comparison and similarity quantification between them \cite{frans_clipdraw_2021}.\par

Our experiments were conducted using stable diffusion models for both generation tasks, adapting the DreamBooth fine-tuning technique with pretrained \textit{stable-diffusion-v1-5} weights as baseline, publicly available in the \textit{Hugging Face} model hub repository \cite{rombach_high-resolution_2022}.

\subsubsection{Healthy image generation} For each dataset we trained a separate model using only healthy images, as each dataset contains independent semantic information in the prompt and because the intensity ranges and image details differ between populations. Additionally, a third model with the combination of mammograms from both vendors was trained, adding to the prompt the vendor's name. \par

We decided not to fine-tune the VAE encoder and decoder after testing its encoding-decoding performance on our mammograms using pretrained natrual images weights. Moreover \cite{chambon_roentgen_2022} found that a pretrained VAE on natural images can perform well on Chest X-ray images. Using this VAE encoder, an original image of 512x512 pixels is compressed to 4 latent representations of 64x64, reducing 16 times its original shape \cite{kingma_auto-encoding_2022}. Consequently, the diffusion process is performed on the latent representations rather than the original images, allowing lower memory usage, fewer layers in the UNet, and faster training and generation.\par

Therefore, only the CLIP text encoder and the UNet weights were trained. The UNet architecture is the original SD UNet proposed by \cite{rombach_high-resolution_2022}. The network has four 2D down- and upsampling blocks. Except for the last downsampling block (and its corresponding upsampling block) all blocks are composed of two ResNet blocks and two transformer blocks, one after the other. The timestep embedding is added to the ResNet blocks whereas the text embedding is added through cross attention into the Transformer blocks. For the last downblock (and first upblock) only the timestep information is fed.\par

The main training hyperparameters explored were the following:

\begin{itemize}
    \item Batch size: 8, 16, 32, 64, 128 and 256.
    \item Training steps: Experiments ranged from 1k up to 16k.
    \item Learning rate: Three main values $1e^{-6}$, $1e^{-5}$, $1e^{-4}$.
\end{itemize}

To select the best hyperparameters and to track the performance of the models, a validation process was conducted by generating 4 sample images from the same random Gaussian noise every $100$ or $200$ training steps. The training loss (mean squared error) and the GPU memory usage were also logged.

\begin{figure}
    \centering
    \includegraphics[width=0.8\textwidth]{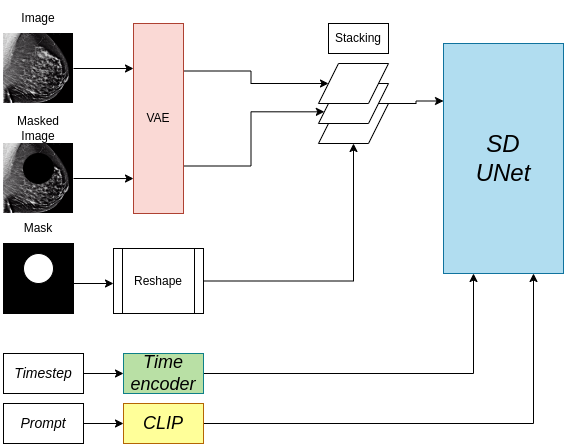}
    \caption{Inpainting training pipeline. The mask is reshaped to match the image size of the latent representations (64x64). The same UNet as in the SD pipeline is used.}
    \label{fig:inpainting_SD}
\end{figure}

\subsubsection{Lesion inpainting} The SD pipeline described for task 1 can be modified in some key aspects to be able to perform the inpainting task. We propose using the modified DreamBooth fine-tuning pipeline to inpaint lesion in a designated region of the breast. \par

For each mammogram with lesion two new elements are added per example: the mask and a masked version of the original image. The masked version means that the pixel values inside of the bounding box are set to zero. At training time, first both the image and the masked image are encoded using the latent space. Also the mask must be reshaped to the latent representation size. The rest of the diffusion process remains the same except for one crucial difference: instead of feeding only the latent representation to the UNet, the latent representation, the mask, and the masked latent representation are stacked into one tensor. This small change in the training process allows the network to pay attention only to the pixels inside the mask, as the pixel outside of it are always provided. This process is described in figure \ref{fig:inpainting_SD}.\par

\section{Experiments}

\subsection{Healthy mammogram generation}

\subsubsection{Independent datasets} 
\begin{figure}
    \centering
    \includegraphics[width=0.9\textwidth]{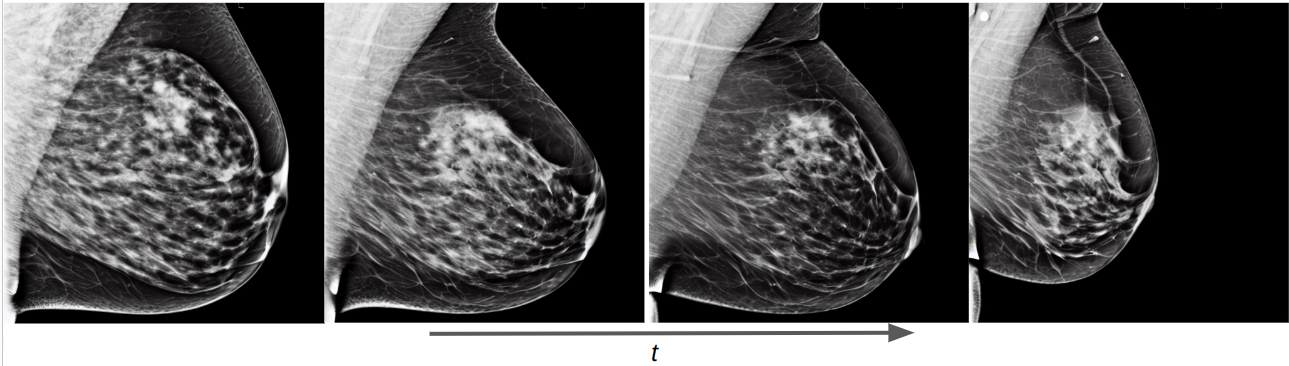}
                \caption{Training evolution of SDM with Hologic images at epoch 1, 3, 6 and 10. The prompt is: "a mammogram in MLO view with small area".}
    \label{fig:conditional_hologic}
\end{figure}

Training examples of the conditional model using prompt text can be shown in figure \ref{fig:conditional_hologic} for the OMI-H dataset. We observe that the fine-tuning technique allows to generate meaningful images since epoch one. For this example we can observe that, as the training process increases, the mammogram reduces its shape in accordance to the area described in the prompt text.\par

Thanks to the combined fine-tuning of the CLIP text encoder and the UNet weights, our conditional models can learn the anatomical structure and form of a mammogram, and can also push the generated image in the direction of the text prompt semantics as the training process increases. \par

\begin{figure}
    \centering
    \includegraphics[width=0.9\textwidth]{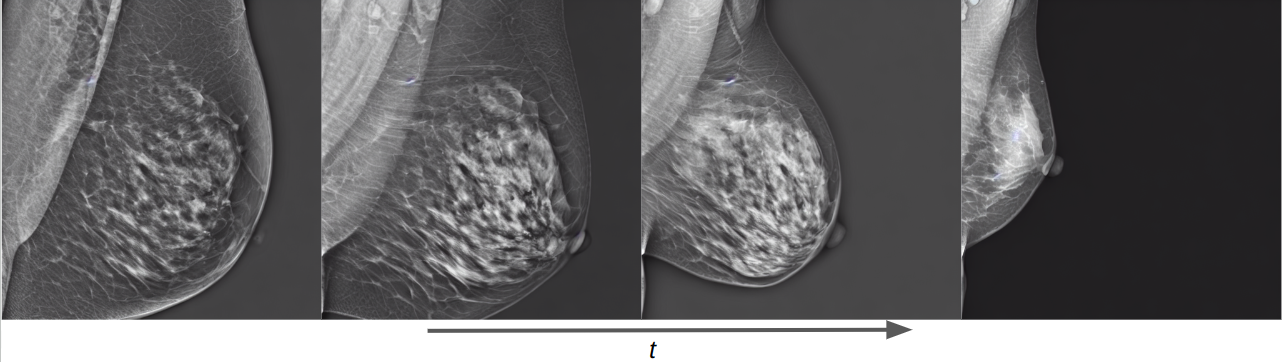}
            \caption{Training evolution of the diffusion process on a conditional pretrained model trained with both Siemens and Hologic images at epoch 1, 3, 7 and 40. The prompt is: "a siemens mammogram in MLO view with high density and small area".}
    \label{fig:conditional_fusion}
\end{figure}

\subsubsection{Concept extrapolation}

Beside allowing us to select the vendor type of the generated mammogram, the combination of both datasets permitted to extrapolate the characteristics of one dataset to the other. This means that, e.g. the breast density of the Hologic mammograms could be controlled, even though this information was not available in the Hologic dataset.

\subsection{Lesion generation}

\begin{figure}
    \centering
    \includegraphics[width=0.99\textwidth]
    {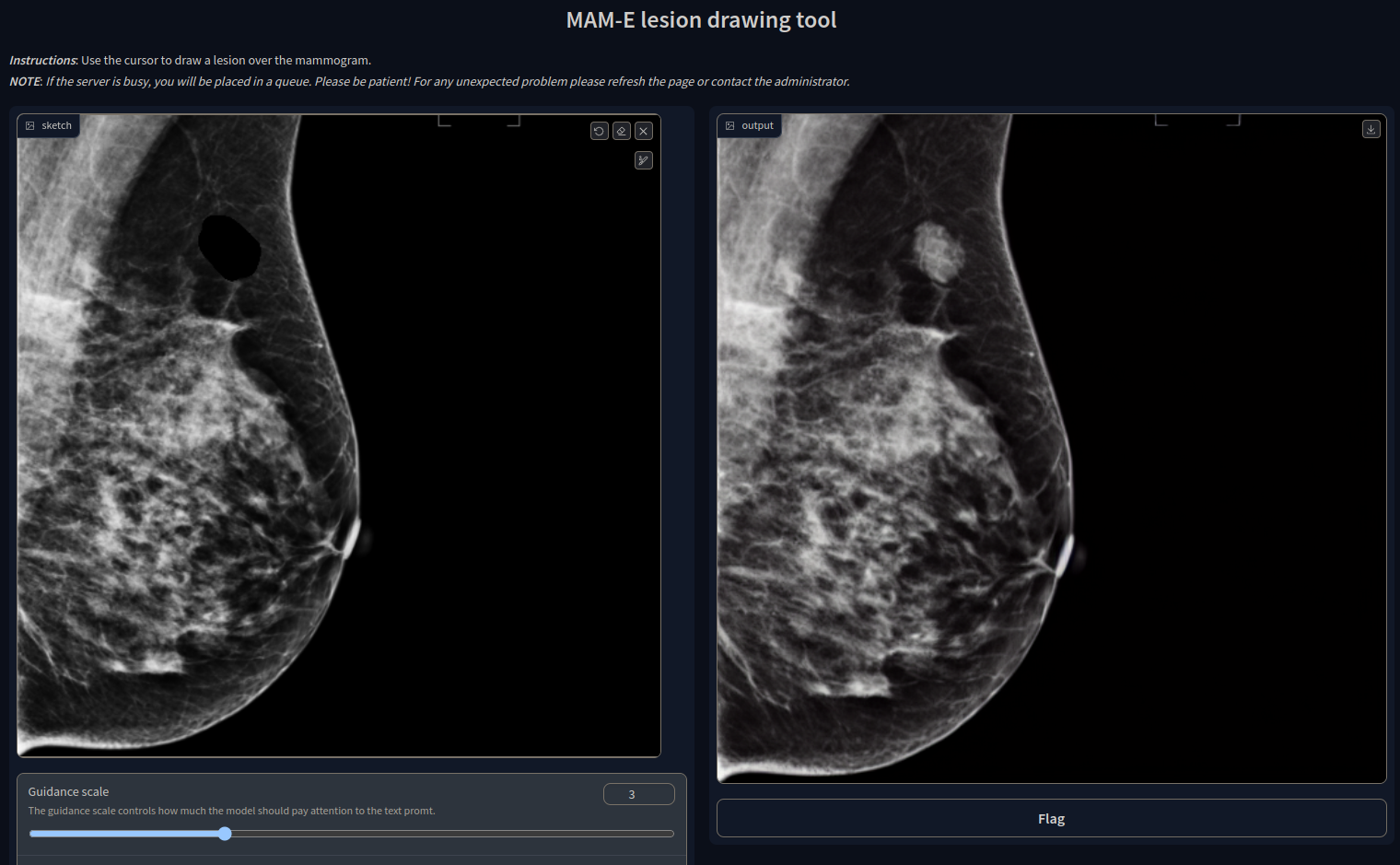}
    \caption{MAM-E lesion drawing tool.}
    \label{fig: mam-e_draw}
\end{figure}

Initial results of the lesion generation pipeline show the possibility to inpaint lesions in any part of the mammogram as shown in figure \ref{fig: mam-e_draw}. As the experimentation with this pipeline was preliminary, only a CAD assessment was conducted to investigate the sensibility of a lesion classification model when a synthetic lesion is presented. Results of this assessment are shown in figure \ref{fig:ROC_and_XAI} and discussed in the following section.

\subsection{Assessment}

\subsubsection{Radiological assessment} 
\begin{figure}[ht]
    \centering
    \includegraphics[width=0.9\textwidth]{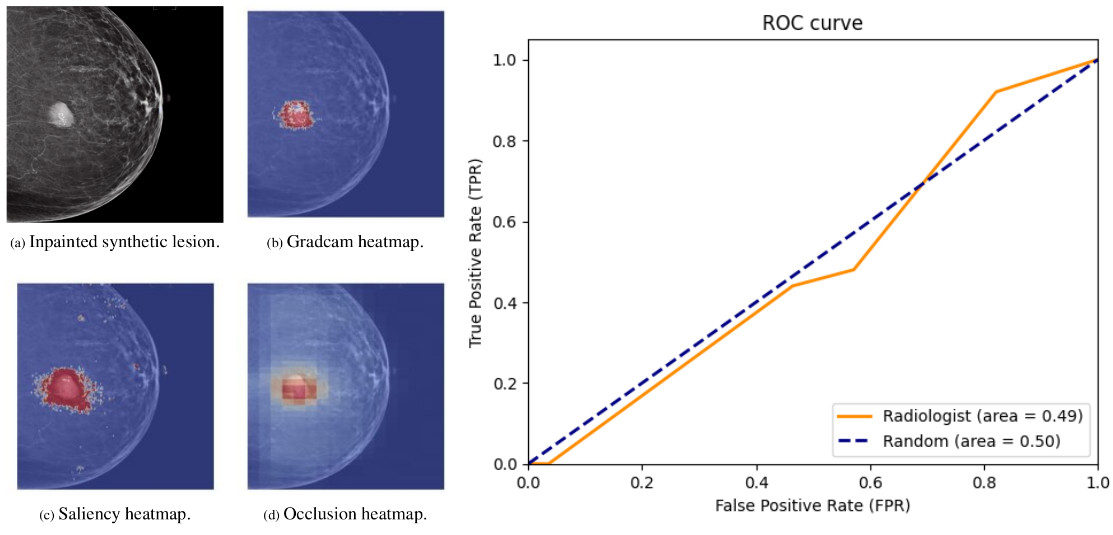}
    \caption{Explainability AI methods heatmaps of synthetic lesion over real healthy mammogram (left) and ROC curve of radiological assessment experiment (right).}
    \label{fig:ROC_and_XAI}
\end{figure}

A visual assessment experiment was performed with the radiological evaluation of 53 synthetic images by a radiologist. The experiment consisted on asking a radiologist with 30 years of experience to rate the mammograms in a scale from 0 (definitely synthetic image) to 4 (definitely real image). The distribution of the mammograms had a 50/50 real-synthetic ratio. The results of the test are summarized as a ROC curve in figure \ref{fig:ROC_and_XAI}. The shape of the ROC curve bears resemblance to the random guess curve, suggesting that the radiologists cannot easily identify the difference between real and synthetic images. Moreover, the AUROC value obtained by the radiologist for this synthetic classification task was 0.49.\par

For the lesion inpainting, the heatmaps of three Explainability AI (XAI) methods were computed for a healthy mammogram with an inpainted synthetic lesion. The CAD system used was a full-field mammogram classification model for benign and malignant breast lesions. The XAI interpretation methods applied were gradcam, saliency and occlusion and their respective heatmaps can be seen in figure \ref{fig:ROC_and_XAI} The hypothesis is that when a synthetic mammogram is used as input the algorithm should highlight the synthetic lesion area, indicating that synthetic lesions have similar pixel distribution to those present in real images.

\section{Conclusion}

Stable diffusion text conditioning is a suitable generative model implementation to synthesize mammograms with specific characteristics and properties. Moreover, fine-tuning a SD model pretrained on natural images with mammographic images is feasible and the training objective is to shift the learned data distribution from a non-medical into our mammography datasets.

We also found that SD can be modified for inpainting of synthetic lesions over healthy mammograms. The developed pipeline essentially only requires the modification of the input latent representation to include a mask to focus the generation process only in that region. All these models inference pipelines were made accessible and ready-to-use through graphical user interfaces, and the weights and code were made available through personal repositories.\par

Thirdly, we found initial evidence that the synthetic images coming from our implementation of SD could potentially be used for CAD systems in need of specific image characteristics or with the presence of lesions. A radiological assessment showed that the initial image quality can be compared with real mammograms and the use of explainability AI models helped to explore the behavior of a classification model.

The first clear limitation of this work is the resolution and pixel depth of the synthetic mammograms. This limited resolution reduces the use of our synthetic images on CAD system that require higher resolution, such as micro-calcification detection. The pixel depth was also reduced from its original 16 bits to 8 bits to match the pretrained model requirements. This reduction losses some information in the images and reduces the overall contrast. With the release of the pretrained weights of SD model for 768x768 resolution images, we expect to perform minimal changes in our current pipeline to allow higher resolution mammography generation. We also plan to train complete CAD pipelines with and without synthetic images to analyze performance changes.

%
%
%
\bibliographystyle{splncs04}
\bibliography{arxiv2023}

\end{document}